\documentstyle[12pt]{article}
\begin{document}
\title{{\hfill{\small{BGU PH-94/03}}\protect\\
\hfill{\small{gr-qc/9408006}}}\protect\\
Finite Magnetic Flux Tube \protect \\
 as a \protect\\
 Black\&White Dihole}
\author{Aharon Davidson\thanks{davidson@bguvms.bgu.ac.il}
 and Edward Gedalin\thanks{gedal@bgumail.bgu.ac.il}}
\date{Department of Physics \\
 Ben-Gurion University of the Negev\\
Beer-Sheva, 84105,  Israel}

\maketitle
\begin{abstract}

A finite-length magnetic vortex line solution
is derived within the context of
(4-dim)  dilaton gravity. We approach the
Bonnor metric at the
Einstein-Maxwell  limit, and encounter the
"flux tube as (Euclidean) Kerr
horizon" at the Kaluza-Klein
level. Exclusively for string
theory, the magnetic flux tube world-sheet
exhibits a 2-dim black\&white
dihole structure.
\end{abstract}
\vfill

\pagebreak

Dilaton gravity is gaining momentum once again.
Several years ago it
was the revival of the Kaluza-Klein (KK) idea which
served to focus attention
on the scalar dilaton ($\equiv$ the local scale of
the extra-dimensional manifold)
couplings induced at the effective 4-dim low energy theory.
At present, the
resumed interest in dilaton gravity is triggered
by string theory,
where dilaton couplings (with a slightly reduced strength
in comparison with KK)
appear rather mandatory when facing the gravitational
consequences of the
theory.
\bigskip

In its simplest version, that is without a scalar potential,
the dilaton
Einstein-Maxwell  action takes  the form

\begin{equation}
\int d^4x\sqrt {-det g}(R + \frac{1}{4}
e^{-2k\eta}g^{\mu\nu}g^{\lambda\sigma}
F_{\mu\lambda}F_{\nu\sigma}+2g^{\mu\nu}\partial_{\mu}
\eta \partial_{\nu}\eta)
\end{equation}

\noindent
Depending on the value of the dilaton coupling constant
$k$, the three special
cases of interest are the following: $k=0$ takes us back
to the Einstein-Maxwell
limit, $k=1$ is suggested by string theory treated to the
lowest order in the
world sheet and string loop expansion, and
$k=\sqrt{1+\frac{2}{N}}$  is
associated  with the $M_4\bigotimes{S_N}$ KK theory.
Of particular interest in
the latter category is the original 5-dim (that is $N=1$)
KK scheme, for
which $k=\sqrt{3}$.

\bigskip

Traditionally,  priority  has  always  been
devoted  to black  hole
solutions ${}^{1-4}$. In this paper, however, we
are a priori after a different
type of solutions (although we recapture black/white
holes as a string
theory bonus), namely axially symmetric solutions which
 represent
finite-length magnetic flux tubes. Being finite, such a
magnetic flux tube must
connect a monopole anti-monopole pair. The work reported
here generalizes in
some respect a preliminary $k=\sqrt{3}$ \ \  KK analysis ${}^{5}$
(for the KK
monopole solution, see ref. ${}^{6}$). We find it therefore
pedagogical to
review first this special case, for which there happens
to exist a simple
(5-dim) geometric interpretation. So, let our starting
point be the 5-metric

\begin{equation}
ds^2_5=- dt^2 + ds^2_{Kerr},
\end{equation}

\noindent
where $-dt^2$ is supplemented by the well known 4-dim
(Euclidean) Kerr metric
$ds^2_{Kerr}$ expressed in the terms of $(x^5,r,\theta,\varphi)$.
To squeeze
out the effective 4-dim theory, one first integrates out the
$x^5$ circle, and
then invokes the Weyl-scaling dimensional reduction procedure
${}^{7}$ $ds^2_5 =
\phi^{-\frac{1}{3}}ds^2_4 + \phi^{\frac{2}{3}}(dx^5 +
A_{\mu}dx^{\mu})^2$.
Following such a standard way of identifying the physical fields
$g_{\mu\nu}(x)$,  $A_{\mu}(x)$, $\phi(x) =exp(-2\sqrt{3}\eta)$,
 and invoking the
explicit  form of the (Euclidean) Kerr metric $ds^2_{Kerr}$,
one recovers the
$k=\sqrt{3}$ limit of our forthcoming eqs.(5-7).

\bigskip

The time is ripe now to present our exact analytic
solution which extremizes
(for arbitrary $k$) the dilaton Einstein-Maxwell action.
Unfortunately, due to
length limitations, we only sketch here some technical
highlights; the detailed
derivation will be presented elsewhere ${}^{8}$. In deriving
this solution we
have closely followed the Chandrasekhar prescription${}^{9}$
of dealing with
stationary, axially symmetric Einstein equations. The reason
is practical.
Although we have a priori content ourselves to a diagonal
metric, the
corresponding equations of motion happen to be the same
as the original Kerr
equations, only with replacements $\Omega$[$\equiv$ Kerr
angular velocity]
$\rightarrow\sqrt{1+k^2} A_{\varphi}$, and
$\psi[\equiv(1/2)\log(-{g_{\varphi\varphi}}/{g_{tt}})]
\rightarrow\psi+2k\eta$. In turn, taking into account the
latter modifications,
one is still led to the complex Ernst equation.
The so-called variable
phase Ernst solution then fixes $A_{\varphi}$ and $\psi
+2k\eta$ combination.
The only new equation in the game

 \begin{equation}
\partial_r(\Delta\partial_r\chi)+(\frac{1}{\sin\theta})
\partial_{\theta}(\delta\frac{1}{\sin\theta}
\partial_{\theta}\chi) =0
\end{equation}

\noindent
is to be solved for $\chi
\equiv(\frac{1}{2})\log(-{g_{\varphi\varphi}}/{g_{tt}})
-(\frac{2}{k})\eta$.
The solution $e^{\chi}=(\Delta \delta)^{\lambda}$ depends
on some constant of
integration $\lambda$ which, at the later stage, must
be chosen as
$\lambda=\frac{1}{2}$ to assure asymptotic flatness.
Here, we use of the familiar
 Boyer-Lindquist notations  (in their Euclidean version)

\begin{equation}
\Delta(r)= r^2-2Mr -a^2,\quad \rho^2(r,\theta) =r^2-
a^2\cos^2\theta,\quad
\delta = \sin^2\theta.
\end{equation}

\bigskip

Collecting the various pieces together, we finally obtain

\begin{eqnarray}
&&ds^2 = (\frac{\Delta +a^2 \delta}{\rho^2})^{\frac{2}{1
+k^2}}[-dt^2 +
\Delta \delta (\frac{\rho^2}{\Delta
+a^2\delta})^{\frac{4}{1+k^2}} d \varphi^2+\\
&&+ (\frac{\rho^2}{\Delta+(M^2+a^2) \delta})^{\frac{3-
k^2}{1+k^2}} \rho^2
(\frac{dr^2}{\Delta} +d\theta^2)],\nonumber\\
&&A_{\varphi}=\frac{1}{\sqrt{1+k^2}}\frac{4aMr
\delta}{\Delta+a^2\delta},\\
&&\eta=\frac{k}{1+k^2}\log{\frac{\rho^2}{\Delta+a^2\delta}}.
\end{eqnarray}

\bigskip

Our general solution is clearly asymptotically-flat.
Moreover, for $M=0$, one
approaches a flat space-time conveniently described by
prolate coordinate
system (where $a$ is recognized as the focal length).
The long distance
behavior of our solution is governed by its mass $m$,
by a magnetic dipole
moment  ${\bf\mu}$, and (for $k\neq0$) by some scalar
 hypercharge as
well. In particular,

\begin{equation}
m=\frac{2M}{1+k^2},\quad{\bf\mu}=\frac{4aM}{\sqrt{1+k^2}}.
\end{equation}

\noindent
For $k=0$, to be referred to as the Einstein-Maxwell
limit, our solution
passes its second consistency check (the first being the
$k=\sqrt{3}$ limit)
by converging to the well known Bonnor metric ${}^{10}$.

\bigskip

Examining
the Riemann tensor, curvature singularities are detected at
$\rho^2=0$ and also along $\Delta  + a^2 \delta = 0$.
As far as $\Delta = 0$ is
concerned, the situation is a bit tricky, and by far
more important for the
sake of our discussion. If $\delta \neq 0$ (that is
$\theta \neq 0,\pi$) the
apparent singularity at $\Delta=0$ turns out to be an
artifact (for arbitrary
$k$), reflecting nothing but an ill-defined coordinate
system. If $\delta = 0$,
on the other hand, and for $k\neq1$ (to be
discussed separately), the so-called end-points
$\theta = 0$ and $\theta = \pi$ of
the $\Delta = 0$ surface forcefully catch the $\Delta +a^2
\delta = 0$  source
singularity!

\bigskip

Approaching the short distance regime, the first surface
 of interest is defined
by the outer zero of $\Delta(r) = 0$, namely

\begin{equation}
r_H = M + \sqrt{M^2+a^2}.
\end {equation}

\noindent
 The geometrical meaning of $r_H$ can be revealed, to
a certain extend, within
the KK embedding ${}^{5}$, see eq. (2), of the $k = \sqrt{3}$
 solution. In such a
5-dim framework, the $r = r_H$ surface is identified as the
Euclidean extension
of the Kerr rotating event horizon, suffering at most
conic singularities
at the $\theta = 0,\pi$ poles. The fact that this surface
(modulo its North and
South poles) stays non-singular even at the 4-dim level, and
this is true for
arbitrary $k$, may indicate a deeper yet quite obscure flux-tube
$\leftrightarrow$ horizon connection.

\bigskip

The crucial observation now is the overall $\Delta$ factor
that $d \varphi^2$ has
picked up. The geometrical significance of this is clear:
as $r \rightarrow
r_H$, the invariant circumferencial length vanishes

\begin{equation}
\int_{0}^{2\pi}\sqrt{g_{\varphi\varphi}} d\varphi \rightarrow 0,
\end{equation}

\noindent
thereby marking the axis of axial symmetry. This is to
be regarded as a
physically non-trivial continuation of a crucial feature
 borrowed from the
underlying  prolate  coordinate system (recall the flat
$M=0$ case, where
prolate $r=a$ is just a line  connecting the two focuses,
whereas $r>a$ are
ellipsoids of revolution).  Altogether, $r = r_H$ defines
 an 1+1-dim (rather
than 2+1-dim) sub-manifold characterized by the following
induced world-sheet
metric

\begin{equation}
ds^2_H = (\frac{a^2\delta}{\rho^2_H})^{\frac{2}{1+k^2}} [-dt^2+
(\frac{\rho^2_H}{(M^2+a^2)\delta})^{\frac{3-k^2}{1+k^2}}
\rho^2_Hd\theta^2],
\end{equation}

\noindent
where $\rho^2_H = r^2_H - a^2 \cos^2\theta$.

\bigskip

The electromagnetic gauge field $A_\varphi(r,\theta)$,
given by eq.(6),
nicely fits into the game. First of all, the $k$ - effect
on $A_\varphi$
(and similarly on $\eta$ ) is remarkably simple, just an
overall $(1 +
k^2)^{-1/2}$ factor. And more important, as $r \rightarrow r_H$

\begin{equation}
\int_{0}^{2\pi} A_{\varphi}d\varphi \rightarrow 4\pi g
\equiv \frac{8 \pi}{\sqrt{1+k^2}}\frac{Mr_H}{a},
\end{equation}

\noindent
indicating magnetic flux confinement. Our $\Delta(r) = 0$
string has become a
magnetic vortex line, stretched between a monopole anti-monopole
 pair of
magnetic charges $\pm g$ at its end-points $\theta = 0$, $\pi$.
 The medium
surrounding the magnetic flux tube is paramagnetically polarized.
 The associated magnetic permeability

\begin{equation}
\exp(2k\eta) =  (1+\frac{2Mr}{\Delta(r)+a^2\sin^2\theta})^{
\frac{2k^2}{1+k^2}}
\end{equation}

\noindent
ranges from unity at spatial infinity up to a blow-up at
the monopole locations.

\bigskip

Given the world - sheet metric eq.(11), the invariant
length $l_k$ of our
vortex string can be calculated by integrating $ds_H$ ( keeping $t$
frozen). $l_k$ gets infinite as $k \rightarrow 0$,
diverging like $k^{-2}$,
thereby presenting a major drawback for the Bonnor dipole .
 The invariant length
is finite, however  for any non-vanishing $k$. Of particular
interest is the
$k=1$ case, representing string theory, for which
$l_1=(2 ar_H/\sqrt{M^2+a^2}) E(a/r_H)$,
 where $E(a/r_H)$
denotes a complete elliptic integral.

\bigskip

Let us now have a closer look at the world-sheet metric
 of our magnetic
flux tube, and  reveal the (2-dim) nature of its monopole
($\theta=0,\pi$)
source singularities. For small $\theta$, that is in the
neighborhood of the North
pole, we reparametrize $\theta \approx \epsilon^{\frac{1
+k^2}{2k^2}}$, to
conveniently  arrive at

\begin{equation}
ds^2_H \approx -\epsilon^{\frac{2}{k^2}}(\omega_k dt)^2
+ d\epsilon^2,
\end{equation}

\noindent
$\omega_k$ being a world-sheet constant easily extracted
from eq. (11). Such an
approximate metric suffices to make our point. Curvature
singularity is
encountered as $\epsilon  \rightarrow 0$, unless $k=1$
 (recall that the Ricci
scalar behaves like $R \approx 2(1-k^2)/k^2\epsilon^2$).
Moreover, the
exclusive feature of the  $k = 1$ two metric (in its
Hawking $t \rightarrow
i\tau$ extension ${}^{11}$) is that the singularity is
only conical, and can
be regarded an artifact provided $\tau$ is identified
with a period of

\begin {equation}
\Delta\tau =\frac{2\pi}{\omega} = 4\pi M(1+\frac{M}{\sqrt{M^2+a^2}}).
\end{equation}

\bigskip

Altogether, we are led to the remarkable conclusion
that $\theta = 0$ acts as a
world-sheet event horizon. A 2-dim hole has emerged.
In fact, there are
\underline{two} of them in the game, as the analysis
for $\theta=\pi$ proceeds
on entirely equal footing. Taking into account local
light-cone considerations,
this may suggest (to be verified soon) the (2-dim)
interpretation of a black
\& white dihole. Hypothetical thermodynamic effects of
the $\tau$-periodicity
lie beyond the scope of the present paper.

\bigskip

Owing to its topological origin, the $\Delta\tau$
periodicity must globally
characterize the flux tube world-sheet. In other words,
the exact $k=1$
two-metric

\begin{equation}
a^{-2}ds^2_H = -\frac{\sin^2\theta}{r^2_H
-a^2 \cos^2 \theta}dt^2 +
\frac{r^2_H-a^2\cos^2\theta}{M^2+a^2} d\theta^2
\end{equation}

\noindent
calls for an analytic Kruskal-type extension ${}^{12}$.
We thus listen to
the $\tau$-periodicity message, and set accordingly
 $u = f(\theta) \cosh \omega
t$, $v = f(\theta) \sinh \omega t$, requiring $f(\theta)$
to be such that
$ds^2_H \propto (du^2-dv^2)$. This algebraic problem
admits two distinct
solutions, they are: $f_N(\theta) = \tan \frac{\theta}{2} \exp
(-\frac{a^2\omega}{\sqrt{M^2+a^2}}\cos \theta)$, for which

\begin{equation}
\frac{\omega^2}{a^2}ds^2_N = \frac{4 \cos^4
\frac{\theta}{2}}{r^2_H-a^2\cos^2\theta}   e^{\frac{2a^2
\omega}{\sqrt{M^2+a^2}}\cos \theta} (du^2_N-dv^2_N),
 \end{equation}

\noindent
and alternatively $f_S(\theta) \equiv f_N(\pi - \theta)$ for which

\begin{equation}
\frac{\omega^2}{a^2}ds^2_S = \frac{4 \sin^4
\frac{\theta}{2}}{r^2_H-a^2 \cos^2 \theta}
 e^ {-\frac{2a^2 \omega}{\sqrt{M^2+a^2}}\cos\theta}
(du^2_S-dv^2_S).
 \end{equation}

\noindent
$f_{N,S}(\theta)$ is regular at the North (South)
 pole, but apparently
troublesome at the opposite pole. A Wu-Yang-like
construction ${}^{13}$ of
matching patches along $\theta=\pi/2$ is quite welcome.

\bigskip

To gain more insight into the structure of local light
cone, a synchronous form
of metric is in order. This can be achieved using the
language of retarded (or
advanced)  Lemaitre coordinates ${}^{14}$ $T = t +
\int\frac{\Gamma}
{1-\Gamma^2} d\xi$ and  $R = t + \int \frac{d\xi}{\Gamma
 (1-\Gamma^2)}$, where
$\Gamma^2(\xi) \equiv \frac{c^2 - 1}{c^2 - \xi^2}$,
 $\xi(T,R)$ is nothing but
the extended $\cos \theta$, and $c = r_H/a \geq1$.
The 2-dim line element
gets transformed into

\begin{equation}
ds^2_H = -dT^2 + \Gamma^2(\xi)dR^2.
\end{equation}

\noindent
which is furthermore free of apparent singularities
and (by being extensible to
include the real singular points) geodesically complete.
The light cone generatrix slope $dT/dR$  varies from
infinity at the $\xi = \pm
c$ singularities, to unity at $\xi = \pm 1$ horizons,
down to minimal $\pm
c^{-1} \sqrt{c^2-1}$ at the central point $\xi =0$.

\bigskip
\begin{figure}[t]
\vspace*{2in}
\caption[1]{(Caption) World--sheet null geodesics in the (retarded) Lemaitre
representation. Notice the different fates of Northerly versus Southerly
directed `light'--rays.}
\end{figure}
The various categories of null geodesics,
illustrated in Fig.1, confirm our black\&white dihole
interpretation. In
particular, notice the different fates of oppositely
directed "light"-rays
through (say) $\zeta=0$. The Northerly directed
ray will unavoidably cross the
black horizon on its way to singularity, whereas
the Southerly directed one
just accumulates on the white horizon ${}^{15}$.
Given the arrow of time, one can
thus tell, on (2-dim) geodesic grounds, a magnetic
monopole  from an anti-monopole. Had we invoked
the advanced (rather
the retarded) Lemaitre
coordinates, we would have faced a rather similar picture,
save for a black
$\leftrightarrow $ white role exchange.

\bigskip

One should wonder by now which of  the above 2-dim
features persists at the
full 4-dim theory. We would like to know whether our
black\&white dihole is
merely a 2-dim reality, or is it in fact a legitimate
4-dim creature? To answer
the latter question in the affirmative, at least locally,
we expand our $k=1$
four-metric around the apparent (say) North singularity
using $r\approx r_H
+\frac{1}{2}\epsilon^2\sqrt{M^2+a^2}$ and $\sin\theta \approx
\theta$. As
advertised,  the result is manifestly flat, free of curvature
singularities, but
with a nontrivial double conic interplay. Namely, for
$\epsilon\ll\theta\ll 1$
we have

\begin{equation}
ds^2\approx ds^2_H + (r^2_H - a^2)[\frac{a^2}{M^2
+ a^2}d\epsilon^2 +\frac{M^2 +
a^2}{a^2} \epsilon^2 d\varphi^2],
\end{equation}

\noindent
whereas $\theta \ll \epsilon \ll 1$ implies

\begin{equation}
ds^2 \approx (r^2_H -a^2)d \Omega ^2 + [-\frac{M^2 + a^2}{r^2_H
-a^2}\epsilon^2 dt^2 +(r^2_H - a^2)d\epsilon^2].
\end{equation}

\noindent
The  good  news  is  that,  independently  on  how  one  approaches
the $(\epsilon,
\theta)$-origin, the \protect\\
imaginary-time conic singularity is
 again removable by
means of the same old periodicity condition eq. (15). This
in turn allows to
maintain the geometrical essence of event horizon. The novel
effect, however,
is the Vilenkin-like cosmic string defect  ${}^{16}$.Once
recognizing the
$\Delta \varphi = 2\pi$ periodicity, on asymptotic and/or
eq. (21)
grounds, a typical  conical structure accompanies eq.(20).

\bigskip

This completes the presentation of a field theoretical
configuration
which can be loosely referred to as a cosmic magnet.
A non-trivial
dilaton coupling is the crucial ingredient needed for
driving the
invariant length finite. On the technical side, we have
recover the
Bonnor  metric at the Einstein-Maxwell limit, and the
"flux tube as a
(Euclidean) Kerr horizon" at the Kaluza-Klein  embedding.
Effective string
theory  gets singled out when analyzing the flux tube
world-sheet geometry. It
is  exclusively for string theory that our flux tube
serendipitously
resembles a (2-dim) black\&white dihole. The latter
physical interpretation,
analytically accepted by the full 4-dim parent theory,
is  supported by a Kruskal
(imaginary time periodic) extension, as well as by  the
Lemaitre (synchronous
light cone and geodesically complete)  representation.

\pagebreak

         {\bf REFERENCES}
\bigskip
\begin{itemize}
\item[1.] G.Gibbons, Nucl. Phys. {\bf B207}, 337 (1982);\protect\\
 G.Gibbons and K.Maeda,
Nucl. Phys. {\bf B298}, 741 (1988).
\item[2.] D.Garfinkle,  G.Horowitz  and  A.Strominger,
Phys.Rev.  {\bf D43},  3140 (1991);\protect\\
D.Garfinkle and A.Strominger, Phys. Lett. {\bf B256}, 146 (1991);\protect\\
J.Horne and G.Horowitz, Nucl.Phys. {\bf B368}, 444 (1992);\protect\\
G.Holzney and F.Wilczek, Nucl.
Phys. {\bf B380}, 447 (1992); \protect\\
T.Banks, M.O'Loughlin and A.Strominger, Phys.Rev.
{\bf D47}, 4476 (1993); \protect\\
R.Gregory and J.Harvey, EFI-92-49 (hep-th/9209070).
 \item[3.] F.Dowker,J.P.Gaunlett, D.A.Kastor and
J.Traschen, EFI-93-51
(hep-th/9309075); \protect\\
F.Dowker, J.P.Gaunlett, S.B.Giddings and G.T.Horowitz,
UCSBTH-93-98 (hep-th/9312172).
 \item[4.] G.Lavrelashvili and D.Maison, Phys.
Lett.{\bf B295}, 67 (1992); \protect\\
 P.Bizon, Phys. Rev. {\bf D47}, 1656 (1993);\protect\\
G.W.Gibbons and C.G.Wells, DAMTP R93/33 (hep-th/9312014).
\item[5.] A.Davidson, "Geometric confinement of
Kaluza-Klein monopoles".
SU-4228-365, 1988 (unpublished)
\item[6.] R.Sorkin, Phys.Rev.Lett. {\bf 51}, 87 (1983);\protect\\
 D.Gross and M.J.Perry, Nucl. Phys. {\bf B226}, 29 (1983).
\item[7.] T.Appelqwist and A.Chodos, Phys.
Rev. {\bf D28}, 772 (1983)
\item[8.] A.Davidson and E.Gedalin (BGU preprint, in
preparation),
\item[9.] S.Chandrasekhar, in "The Mathematical Theory of Black
Holes" (Oxford Univ.Press, 1983)
\item[10.] W,B.Bonnor, Z.Phys. {\bf 190}, 444 (1966);
 J.Phys. {\bf A12}, 853 (1979)
\item[11.]  S.W.Hawking, Phys. Rev. {\bf D13}, 191 (1976).
\item[12.] M.D.Kruskal, Phys.Rev. {\bf 119}, 1743 (1960);\protect\\
 P.Szekeres, Publ. Math. Deb. {\bf7}, 285 (1960).
\item[13.] T.T.Wu, C.N.Yang, Phys. Rev. Lett.{\bf 13}, 380 (1964).
\item[14.] G.Lemaitre, Ann.Soc.Sci.Bruxelles,Ser.{\bf A53}, 5
(1933);\protect\\
 L.Landau and E .Lifshiz, in  "The Classical
Theory of Fields", sec.102
(Pergamon, London 1975).
\item[15.]I.Q.Novikov, Astro. J. (in Russian){\bf 41}, 1075
(1964);\protect\\
 Y.Ne'eman, Astro.J.{\bf141}, 1303 (1965).
\item[16]. A.Vilenkin, Phys. Rev. {\bf D23}, 852 (1981).
 \end{itemize}
\end{document}